# Large-scale Fabrication of High Density Silicon-vacancy Centers via Helium-ion Implantation of Diamond Nucleation Surface

*Chengyuan Yang, Zhaohong Mi, Huining Jin, and Andrew Bettiol\**

Department of Physics, National University of Singapore, 2 Science Dr 3, 117542, Singapore

ABSTRACT

Silicon-vacancy (SiV) color centers in diamond have great potential for optical sensing and bio-imaging applications. However, the fabrication of large-scale high density SiV centers in diamond remains difficult. Here, we report a promising method for the fabrication of high density $SiV^-$ centers in a low-cost polycrystalline diamond film grown on an inches-scale Si wafer. Our method utilizes the nucleation surface of the diamond film which initially interfaces with the Si wafer. Benefited from the diamond seeding substrate of silicon, the nucleation surface has originally incorporated with high density Si atoms. Upon helium-ion implantation and subsequent thermal annealing, we demonstrate by performing PL mapping that these Si atoms can be efficiently converted to SiV¯ centers. These SiV¯ centers exhibit bright emission and a relatively long fluorescence lifetime (~1.08 ns) that is comparable to the SiV¯ lifetime reported in single crystal diamond. Furthermore, by using a focused helium beam and varying the helium fluence, we demonstrate the feasible density control and patterning of the SiV¯ centers. These results show that our method can produce high density SiV¯ centers in low-cost wafer-scale polycrystalline diamond, which could facilitate the commercialization of SiV¯ centers-based optical devices.

# 1. Introduction

Optically active point defects in diamonds have gained increasing attention in the fields of quantum computing,[1] optical sensing,[2] and biological imaging.[3] Among these defect centers, negatively charged silicon-vacancy (SiV¯) centers exhibit a narrow spectral bandwidth (~0.7 nm) at room temperature,[4] high brightness (emission rate ~$6\times10^6$ s$^{-1}$),[5] and a short excited-state lifetime (~1 ns).[6–8] Moreover, the SiV¯ center enables feasible optical initialization and readout of its electronic spin.[9] Such properties make SiV¯ attractive as quantum emitters,[10] biological probes, and magnetic[11] and thermal[12] sensors.

Although SiV¯ center engineering in single crystal diamond has been routinely realized,[13] single-crystal diamond is essentially limited by the sizes achievable. Despite some costly methods were attempted to develop large diamond plates,[14] it remains challenging for cost-effective and large-scale (up to inches) fabrication of SiV¯ centers in a single-crystal diamond, especially, for industrial development. Alternatively, polycrystalline diamond films can be used for large scale hosting of SiV¯ centers. Although higher in optical loss, polycrystalline diamond films offer advantages of low-cost and large-scale (more than 8 inches). Furthermore, polycrystalline diamond films can be grown on Si wafers, which enables Si incorporation through the growth process. As it has been reported,[15] a Si concentration up to an order of $1\times10^{19}$ cm$^{-3}$ can be achieved in the nucleation surface vicinity where the diamond is seeded on the Si. The high density Si atoms offers a sufficient Si source for fabricating the SiV¯ centers and also eliminates the need for Si-doping processes such as by Si ion implantation.[16–18]

A challenge for the polycrystalline diamond lies in efficiently converting the Si atoms to the SiV¯, as most of the Si atoms were incorporated at substitutional sites of the diamond lattice.[19] To efficiently generate SiV¯ centers, creating lattice vacancies and subsequently annealing the diamond to combine the vacancies with the Si atoms can be a solution. Among the ion species (electron, proton, etc) for creating diamond vacancies, helium-ions have shown the highest efficiency.[20] Furthermore, as a type of light ions, helium ions exhibits much lower radiation damage than heavy ions, therefore the color centers suffer less fluorescence quenching and are able to maintain high brightness and long coherent time, as proved by nitrogen-vacancy centers.[21]

Here, we report large-scale fabrication of dense SiV¯ centers in the polycrystalline diamond based on performing helium-ion implantation at the diamond nucleation surface. In our approach, commercial polycrystalline diamond films grown onto Si wafers were employed. After being delaminated from the Si wafer, the films were implanted with mega-electron-volts (MeV) helium ions at the nucleation surface and subsequently annealed in a vacuum environment. We show that a thin layer (about 750 nm) of bright and uniformly distributed SiV¯ centers are formed in the region of the shallow nucleation surface of the diamond film. By focusing the helium-ion beam to a sub-µm spot size and varying the ion fluences, we demonstrate the feasible density control and patterning of the SiV¯ centers on the nucleation surface. Considering the affordability and wafer-scale size of the polycrystalline diamond films, our method offers a promising route for large-scale fabrication of SiV¯ centers in diamond.

## 2. Experiment method

The polycrystalline diamond films (thickness about 10 µm), grown on 6-inch Si substrates, were supplied by Blue Wave Semiconductor, Inc. Figure 1a shows a schematic diagram of the fabrication process. To get the nucleation surface exposed, the diamond film was first delaminated from the Si substrate by mechanic peeling. Then the nucleation surface of the diamond film was implanted with 1.7 MeV helium ions which were produced by a Singletron ion accelerator (High Voltage Engineering Europa B.V.). Lastly, the diamond film was annealed in a furnace at 800 °C for 2 hours at a vacuum condition ($< 1\times10^{-5}$ mbar).

Optical characterization of the diamond film was carried out at room temperature by a confocal microscope (a pinhole size of 50 µm) equipped with an Andor SR-500 optical spectrometer. The spectral resolution is 0.7 nm (Grating 150 l/mm) and 0.12 nm (Grating 1200 l/mm) for photoluminescence (PL) and Raman spectroscopy, respectively. A 532 nm solid state CW laser focused by a 100× objective lens (Olympus UMPLANFI, NA 0.95) was used for PL mapping and spectroscopy. Fluorescence lifetime was measured by exciting the SiV¯ centers with a 532nm pulsed laser generated from a super continuum laser (pulse width of 800 ps, repetition rate 76 MHz). The fluorescence from SiV¯ centers was detected by an avalanche photodetector (APD model: SPCM-AQRH-14-FC) which was synchronized with the laser by a time correlated single photon counter (TimeHarp 260, Picoquant).

# 3. Results and discussion

### 3.1 Fabrication of SiV⁻ centers

Figure 1b shows a piece of a delaminated polycrystalline diamond film with a size about 25 x 7 mm$^2$. Figure 1c shows that the film is flexible, which is bendable from a tangential angle of +9° to a negative angle of -3°. Bowing of the film is due to a gradient in stress along the growth direction.[22] The thickness of the film was measured to be 10±0.5 μm with an optical microscope (Figure 1d).

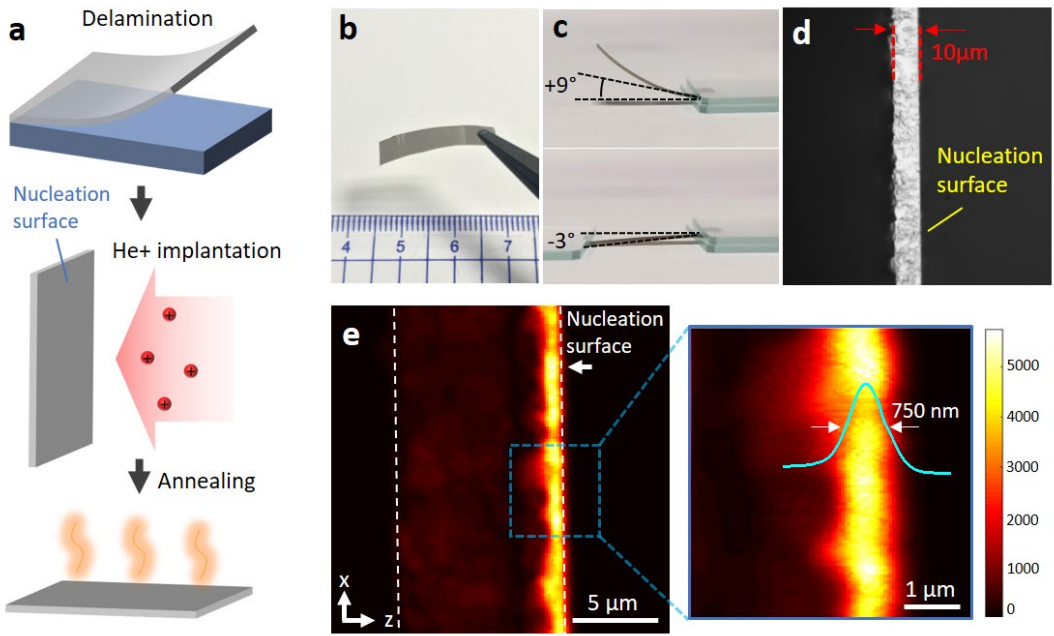

**Figure 1.** Fabrication and optical characterization of the SiV⁻ centers in the diamond film. (a) Fabrication process: the diamond film is first delaminated from the Si substrate, and then implanted with helium ions at the nucleation surface, then annealed in a vacuum environment. (b) Photo showing the delaminated polycrystalline diamond film, together with the illustration of its sizes. (c) Demonstration of the flexibility of the diamond film. (d) Cross-sectional optical image of the film showing its thickness and the edge of the nucleation surface. (e) A 20×20 μm2 cross-sectional PL map showing the emission of the SiV⁻ centers in the diamond film. Inset shows a depth profile of the SiV⁻ centers. A spectral window of 735 to 741 nm was used for mapping the SiV⁻ emission.

After the helium ion implantation (fluence of $2\times10^{15}$ cm$^{-2}$) and the annealing processes, the diamond film was characterized with photoluminescence (PL) spectroscopy. By performing PL mapping on the cross-sectional surface of the diamond film, we observed strong emission within a thin layer close to the shallow nucleation surface (Figure 1e). The emission was confirmed to source from SiV¯ centers by the PL spectrum, identified by the zero-phonon line (ZPL) of 738 nm (Figure 2a). Compared to a single SiV¯ center (linewidth of 0.7 nm[4]), our spectral linewidth is broadened to ~7 nm. We attribute such a broadening to the mechanical stress in the diamond lattice that slightly shifts of the PL spectrum of each SiV¯ center in the light collection area.[23] Depth profiling of the SiV¯ centers (inset in Figure 1e) shows the SiV¯ centers were distributed in the vicinity of the nucleation surface, from the surface to a depth of about 750 nm, as estimated by the full width half maximum (FWHM) of the emitting layer. The SiV¯ layer is uniform, given by the PL intensity along the nucleation surface which shows a relative standard deviation of only 8.4%.

### 3.2 Effect of the helium-ion implantation

The effect of the helium-ion implantation was investigated. Figure 2a shows the PL spectra of the processed film and the as-grown film. We can see that, upon helium-ion implantation and subsequent annealing of the diamond film, the processed film shows a significantly higher SiV¯ emission compared to the as-grown sample. Note that the Si concentration in the film remains the same before and after the ion implantation and annealing processes, since the annealing temperature (800 °C) is unable to mobilize the Si atoms.[24] Therefore, the stronger emission from the processed sample was a result of an efficient conversion of Si to SiV¯ centers in the diamond film. Moreover, Figure 2a also shows that sole annealing of the as-grown sample did not result in a significant increase of the SiV⁻ emission. Direct PL intensity comparison between the helium ion implanted area and un-implanted area, as shown in Figure 2b, indicates that the helium ion implantation is critical for an efficient generation of the SiV⁻ centers.

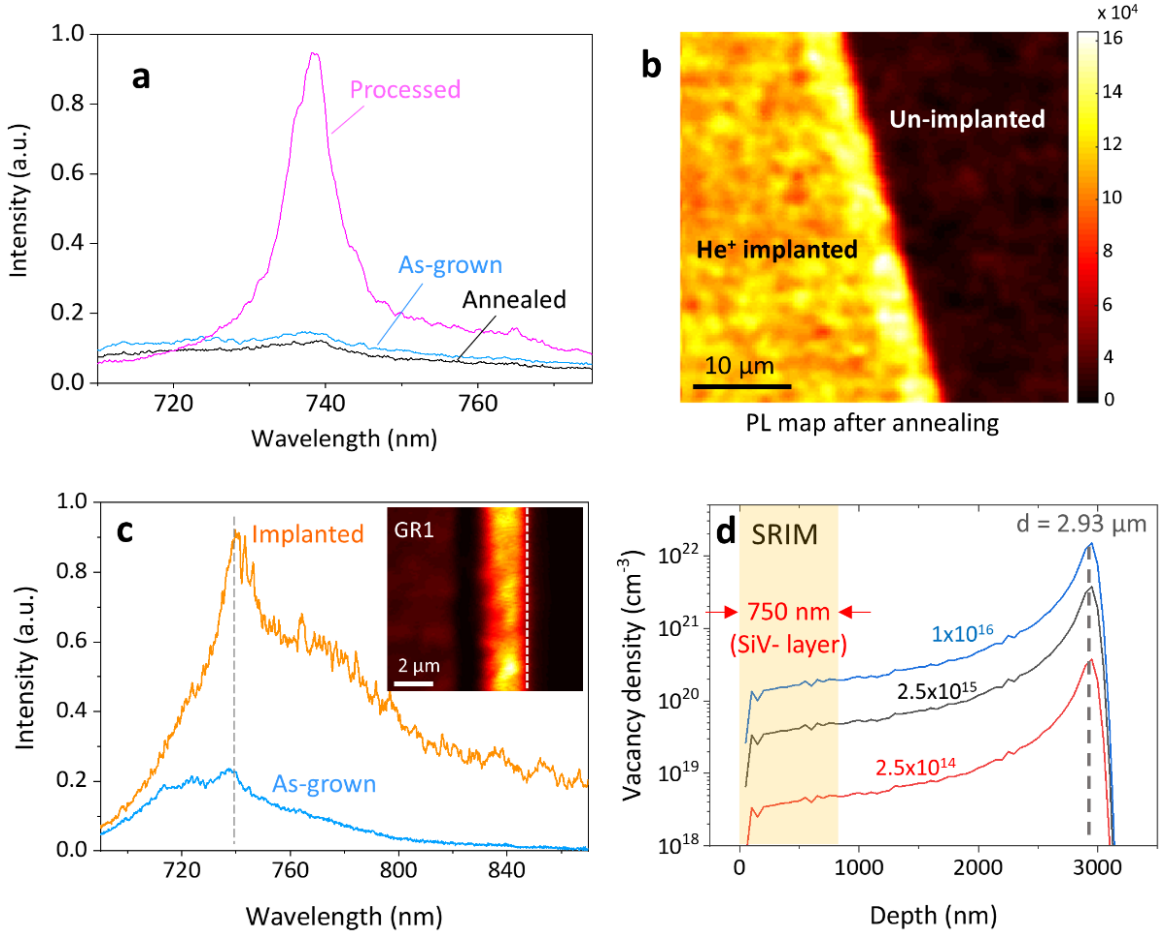

**Figure 2.** PL characterization of the treated and the pristine diamond films. (a) PL spectra of the diamond film of the as-grown, the only annealed, and the processed sample with both the helium-ion implantation and annealing processes. (b) A 40×40 μm² PL map of the SiV- centers on the nucleation surface after annealing. The bright area has been implanted with helium ions (fluence of $2\times10^{15}$ cm$^{-2}$). A spectral window of 735 to 741 nm was used. (c) PL spectrum taken after helium-ion implantation but before annealing ('Implanted'), in contrast to the as-grown film. The dashed line on the spectra indicates the wavelength at 740 nm. Inset shows the cross-sectional PL map of the GR1 centers in the spectral window of 740 to 850 nm. (d) Vacancy density distribution along with the depth for ion fluences of $2.5\times10^{14}$, $2.5\times10^{15}$ and $1\times10^{16}$ cm$^{-2}$, calculated by SRIM[25]. A diamond density of 3.51 g/cm³ and a displacement energy of 43.37 eV (averaged over [100], [111] and [110] directions[26]) are used.

Lattice vacancies created by the helium ions were investigated by the PL spectrum of the diamond film after the helium-ion implantation (without annealing), shown in Figure 2c. Compared to the as-grown film, the helium-ion-implanted film showed a broadband emission from single neutral vacancies (GR1 centers), as identified by the ZPL of 740 nm. The GR1 emission indicates single vacancies were created in the nucleation layer by the helium ions. Cross-sectional PL map of the GR1 centers (inset in Figure 2c) shows the single vacancies were distributed from the surface to a depth of about 2.3 µm, which is consistent with the penetration depth of the helium ions (2.9 µm, calculated by SRIM[25]). The shallower depth of the single vacancies than the ion penetration depth is due to the formation of vacancy clusters at the end-of-range of the ions.

To investigate the influence of the density of the single vacancies on the SiV¯ centers, a series of helium ion fluences ranging from $2.5\times10^{14}$ cm$^{-2}$ to $1\times10^{16}$ cm$^{-2}$ were used. The corresponding vacancy density in the nucleation layer where SiV¯ centers are formed (the SiV¯ layer in Figure 2d) was calculated in an order of $10^{18}$ to $10^{20}$ cm$^{-3}$, which is far below the atomic density of diamond crystal ($1.76\times10^{23}$ cm$^{-3}$). Therefore, negligible fluorescence quenching caused by radiation damage is assumed. This enables us to infer the density of the color centers from their PL intensity, that is, the higher the PL intensity, the higher density of the color center. To correlate the density of the SiV¯ centers with the single vacancies, PL intensities of the GR1 centers and the SiV¯ centers were measured from their corresponding PL spectra of the nucleation surface before and after the annealing process (Figure 3a), and plotted as a function of the helium-ion fluence (Figure 3b).

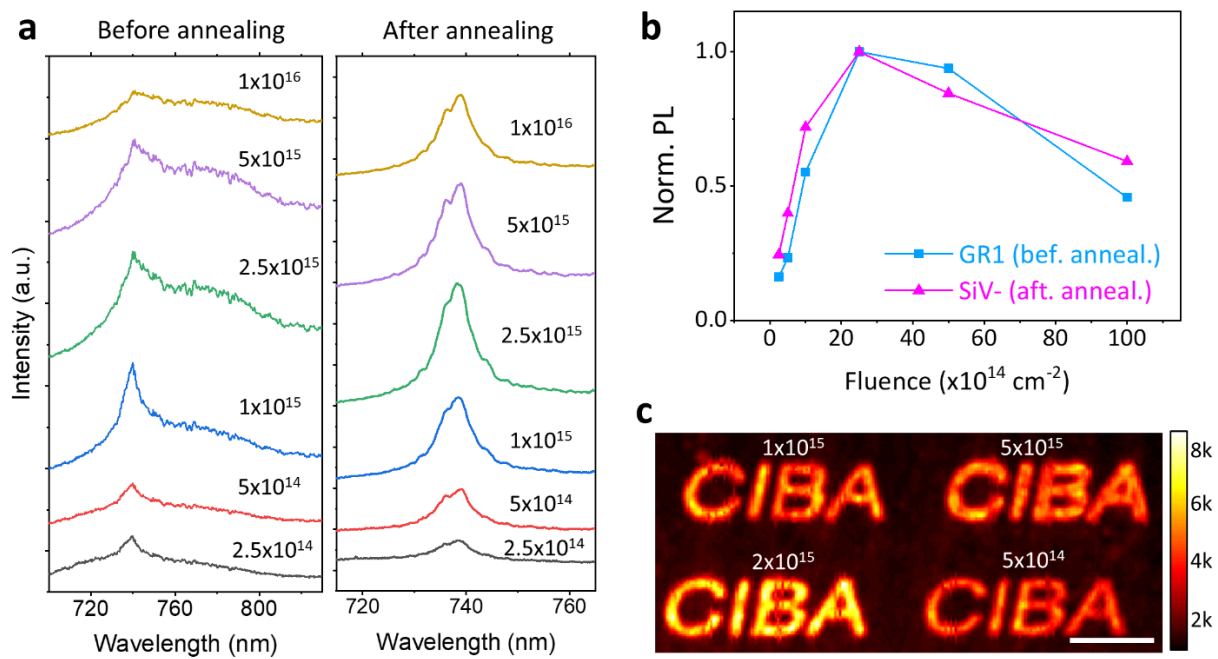

**Figure 3.** Helium-ion implantation and thermal annealing study. (a) PL spectra of the nucleation surface before and after the annealing at the temperature of 800°C. A series of ion fluences ranging from $2.5\times10^{14}$ cm$^{-2}$ to $1\times10^{16}$ cm$^{-2}$ were used. (b) Normalized PL intensities of the GR1 centers (before annealing) and the SiV$^-$ centers (after annealing) at different ion fluences. The maximum PL intensity at the optimal fluence of $2.5\times10^{15}$ cm$^{-2}$ was normalized to unity for both the GR1 centers and the SiV$^-$ centers. PL intensities of the GR1 centers and the SiV$^-$ centers were calculated by integrating the corresponding PL spectrum in the window of 740 to 820 nm and 730 to 750 nm, respectively. (c) PL map of patterned SiV$^-$ centers on the nucleation surface. The map size is 60×30 µm2. Fluences were varied from $5\times10^{14}$ to $5\times10^{15}$ cm$^{-2}$. Scale bar: 10 µm.

Figure 3b shows that the SiV$^-$ emission follows the same trend as the GR1 emission, implying a strong dependence of the formation of the SiV$^-$ centers on the density of single vacancies. This can be explained from the fact that a Si atom needs capturing a vacancy to form a SiV$^-$ center. Before the helium ion implantation, the diamond film has sufficient Si atoms, but most of them are at substitutional sites. The helium-ion implantation creates single vacancies in the diamond film, so initially the GR1 emission increases with the helium ion fluence. As more single vacancies are created, the chance for a Si atom capturing a vacancy during the annealing process increases, leading to an increased SiV¯ density with the helium ion fluence. As the fluence further increases, exceeding $2.5\times10^{15}$ cm$^{-2}$, vacancy clusters are increasingly formed, as we can infer from the decreased emission of GR1 centers. The decrease of the density of single vacancies lowers the conversion efficiency of the Si to SiV¯ centers. Therefore, the SiV¯ emission decreases in the high fluence range. The strong dependence of the SiV¯ density on the helium-ion fluence can be utilized for patterning SiV¯ centers. By using a focused helium ion beam, we demonstrated in Figure 3c that SiV¯ patterns with different densities can be fabricated in the nucleation surface.

By performing SRIM simulation, we estimated the SiV¯ density in the diamond nucleation layer. At the optimal helium-ion fluence of $2.5\times10^{15}$ cm$^{-2}$, the calculated vacancy density is about $4\times10^{19}$ cm$^{-3}$ in the SiV¯ layer. By assuming that the majority of the vacancies are single vacancies and the chance for a Si atom capturing a vacancy to form SiV¯ center is 20% (referring to an E-beam irradiated diamond[27]), we estimated the SiV¯ density in the nucleation layer is of the order about $10^{18}$ cm$^{-3}$, which is remarkable considering that such dense SiV¯ centers are generated in a commercial polycrystalline diamond film without additional Si ion implantation or in-situ doping.

### 3.3 Fluorescence lifetime of the SiV¯ centers

Performance of the SiV¯ centers was evaluated by their fluorescence lifetime. Figure 4a shows the PL decay of the SiV¯ centers excited by a 532 nm pulsed laser. Deconvolution was carried out to compensate the delay from the excitation laser. By fitting the deconvoluted PL decay with a single exponential decay function (inset in Figure 4a), we obtained a fluorescence lifetime of 1.08 ± 0.1 ns for our SiV¯ centers.

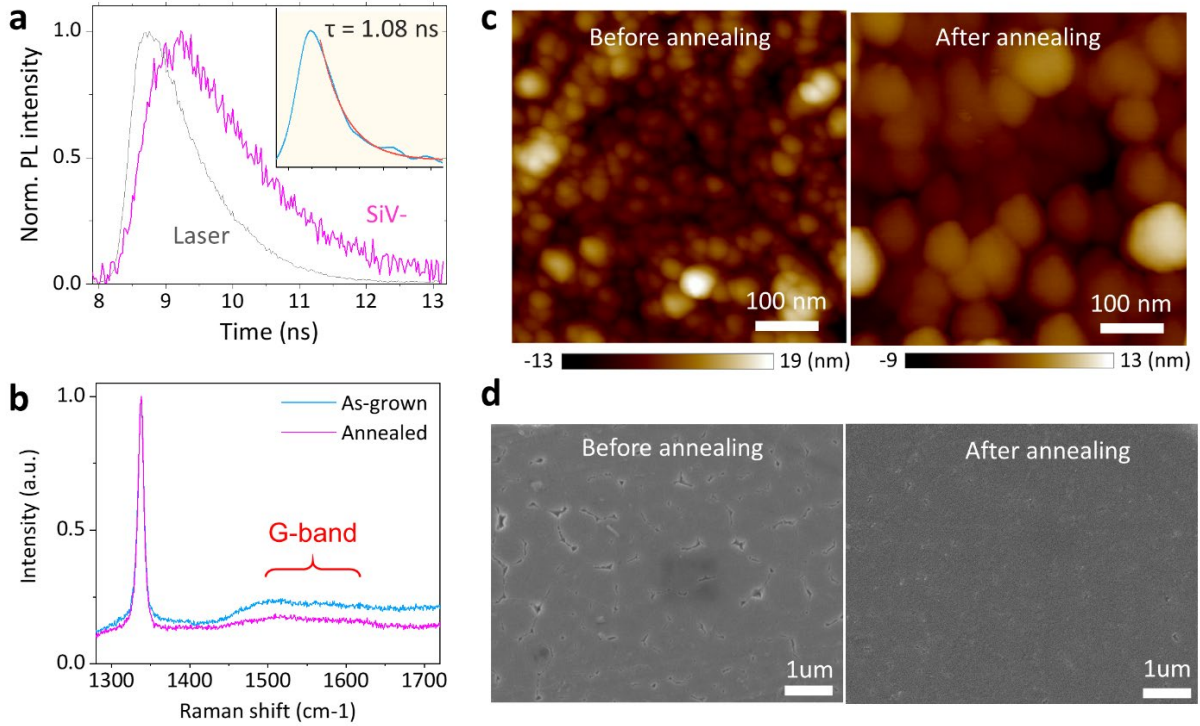

**Figure 4.** (a) PL decay of the SiV¯ centers at room temperature showing a lifetime of 1.08 ns. The SiV¯ centers were fabricated at the helium ion fluence of $2.5\times10^{15}$ cm$^{-2}$. Inset is the PL decay after deconvolution. (b) Raman spectra of the diamond nucleation layer before ("As-grown") and after annealing ("Annealed"). To compare the G-band intensity of the annealed and the as-grown films, the diamond Raman peak (at 1332 cm$^{-1}$) was normalized to unity. (c) AFM images, and (d) SEM images of the nucleation surface taken from the same diamond film before and after the annealing process.

Fluorescence quenching from non-radiative defects at grain boundaries often cause a shorter fluorescence lifetime of color centers in polycrystalline diamond, compared to that in single-crystal diamonds. However, our result shows a comparable lifetime to that in single crystalline diamonds (1.03 to 1.41 ns at 300 K[6–8]). By performing Raman spectroscopy (Figure 4b), we found that our long SiV¯ lifetime results from a significant removal of amorphous carbons by the annealing process, as shown by the lower G-band (1500-1600 cm$^{-1}$) of the annealed film compared to the as-grown film. From the AFM images of the nucleation surface (Figure 4c), we found that the diamond grains coalesce during the annealing process, resulting in a significant increase of the grain size from 20-60 nm to about 100 nm in diameter. As the surface-to-volume ratio is inversely proportional to the diameter of the grain, a larger grain size leads to less surface area for grain boundaries per unit volume, therefore, less amorphous carbons in the annealed film. Furthermore, SEM images (Figure 4d) indicate that the coalescence during the annealing process can also smooth surface cracks on the nucleation surface. The averaged roughness is reduced from 5.7 ± 1 nm to 5.2 ± 1 nm after annealing, measured by AFM in a 5×5 μm$^2$ area.

## 4. Conclusion

In conclusion, we have demonstrated large-scale fabrication of dense (up to the order of 10$^{18}$ cm$^{-3}$) SiV$^-$ centers in a commercial polycrystalline diamond film through helium-ion implantation at the nucleation surface. By varying the helium ion fluence, we identified that the density of single vacancies is an important factor for the conversion efficiency of the Si atoms to SiV$^-$ centers. The optimal helium ion fluence for producing the highest SiV$^-$ density is 2.5×10$^{15}$ cm$^{-2}$. Moreover, minimal fluorescence quenching to the SiV$^-$ centers was observed, as was evidenced by the relatively long fluorescence lifetime (1.08 ns at room temperature). Controlled patterning of SiV$^-$ centers was also demonstrated by using a focused helium ion beam, which can be further exploited for deterministic fabrication of SiV$^-$ centers. Benefited from the high density of the SiV$^-$ centers and the excellent mechanic properties and scalability of the diamond film, our approach will open a new path to building robust and flexible diamond devices for large-scale applications such as optical sensing and bio-imaging.

**Corresponding Author**

*Email: a.bettiol@nus.edu.sg

**Present Addresses**

† Current address of H.J. is Shanghai Microelectronics Equipment Co., Ltd at No.1525 Zhangdong Road Zhangjiang Hi-Tech Park Shanghai, China.

**Author Contributions**

A.B. conceived the work. C.Y. designed the experiment. Z.M. contributed to focused helium ion beam implantation. C.Y. conducted the optical characterization. All authors contributed to writing and editing of the manuscript.

ACKNOWLEDGMENT

We would like to acknowledge Blue Wave Semiconductor, Inc for providing the diamond wafers.